\documentclass[smallextended]{svjour3}       
\smartqed  
\usepackage{graphicx}
\usepackage{booktabs}
\usepackage{xspace}
\usepackage[hidelinks]{hyperref}
\usepackage{jlcode}

 \journalname{Computing and Software for Big Science}
 
\newcommand{\julia}{\textsc{Julia}\xspace}
\newcommand{\jupyter}{\textsc{Jupyter}\xspace}
\newcommand{\python}{\textsc{Python}\xspace}

\newcommand{\numpy}{\textsc{NumPy}\xspace}
\newcommand{\scipy}{\textsc{SciPy}\xspace}
\newcommand{\cplusplus}{\textsc{C++}\xspace}
\newcommand{\fortran}{\textsc{Fortran}\xspace} 
\newcommand{\go}{\textsc{Go}\xspace} 
\newcommand{\geant}{\textsc{Geant4}\xspace}

\newcommand{\lcio}{\textsc{LCIO}\xspace}
\newcommand{\cernroot}{\textsc{Root}\xspace}
\newcommand{\pyroot}{\textsc{PyRoot}\xspace}
\newcommand{\pythia}{\textsc{Pythia}\xspace}
\newcommand{\whizard}{\textsc{Whizard}\xspace}
\newcommand{\pandorapfa}{\textsc{PandoraPFA}\xspace}
\newcommand{\fastjet}{\textsc{FastJet}\xspace}
\newcommand{\sid}{\textsc{SiD}\xspace}
\newcommand{\epem}{\mbox{${\rm e}^+{\rm e}^-$}\xspace}
\newcommand{\mpmm}{\mbox{${\rm \mu}^+{\rm \mu}^-$}\xspace}

\makeatletter  
\providecommand*{\toclevel@lstlisting}{1}  
\makeatother 
 
\begin{document}

\title{Performance of Julia for High Energy Physics Analyses
}


\author{Marcel Stanitzki \and Jan Strube
}

\institute{Marcel Stanitzki \at
              DESY \\
              Notkestrasse 85, 22607 Hamburg, Germany \\
              Tel.: +49 40 89984930 \\
              \email{marcel.stanitzki@desy.de}
           \and
           Jan Strube \at
              University of Oregon\\
              1371 E 13th Avenue, Eugene, OR 97403 \\
              and \\
              Pacific Northwest National Laboratory \\ 
              902 Battelle Blvd, Richland, WA 99352, USA \\
              Tel.: +1 509.375.2217 \\
               \email{jan.strube@pnnl.gov}
}

\date{Received: date / Accepted: date}

\maketitle

\begin{abstract}
We argue that the \julia programming language is a compelling alternative to implementations in \python and \cplusplus 
for common data analysis workflows in high energy physics. 
We compare the speed of implementations of different workflows in \julia with those in \python and \cplusplus. Our studies show that the 
\julia implementations are competitive for tasks that are dominated by computational load rather than data access. For work that is dominated 
by data access, we demonstrate an application with concurrent file reading and parallel data processing.
\keywords{High Energy Physics \and Julia Programming Language \and Data Analysis \and Multithreading}
\end{abstract}

\section{Introduction}
\label{section:introduction}
In high energy physics, the programming language of choice for most of the large-scale experiments, like BaBar and Belle, CDF and D0, and the LHC experiments, has been \cplusplus. It is used for the simulation and reconstruction chain, 
 as well as for data analysis, which also makes heavy use of the \cernroot~\cite{ANTCHEVA20092499} data analysis framework.
Recently, a change in this paradigm has been observed, where data analysis has been moving towards using \python.
Reasons for this move include the faster turn-around of a dynamic scripting language, a lower hurdle for beginners and the growing
versatility of \python thanks to packages like \numpy~\cite{oliphant2006guide} and \scipy~\cite{2020SciPy-NMeth}, developments 
like the \python--\cplusplus bindings for \cernroot (\pyroot~\cite{fons_rademakers_2019_3895860}), and data science packages like 
\textsc{pandas}~\cite{pandas}. Most of the industry-standard deep learning tool kits include \python bindings.

A frequent criticism of \python compared to compiled languages like \cplusplus or \fortran is 
the performance penalty due to the dynamic character. The \julia programming language~\cite{doi:10.1137/141000671} originates from the high-performance 
numerical analysis community and is designed to combine the benefits of dynamic languages like \python 
with the performance of compiled code by using a just-in-time compiler (JIT) approach.
In this paper we demonstrate that \julia exhibits three key features that make it uniquely suitable for data analysis in high energy physics:
\begin{enumerate}
        \item \julia is fast: We perform benchmarks of data analysis tasks that are typical in high energy physics, implemented in 
        \julia, \python and \cplusplus and compare their performance.
        \item \julia is easy to interface with existing \cplusplus libraries: As an example, we have created an interface to the 
        \fastjet library~\cite{Cacciari:2011ma}, the de-facto standard implementation of jet clustering in high energy physics.
        \item \julia has support for parallelism and concurrency: We demonstrate how even a trivial implementation of event-level parallelism can accelerate 
         realistic analysis workflows, even if the underlying I/O libraries do not support multi-threading.
\end{enumerate}
Additionally, \julia is interactive, being one of the original languages that integrates with \jupyter~\cite{Kluyver:2016aa,Jupyter} 
notebooks.

The first public release of \julia code was on February 13$\mathrm{th}$, 2013. Within high energy physics (HEP), \julia has been met with interest early
on~\cite{rootws2015:pata:julia,dianahep2016}, even before the release of 1.0 on August 8, 2018, with its promise of API stability throughout the release 1.x cycle. 
We argue that adoption by a wider user base was for a while largely hampered by the lack of readers of HEP-specific file formats, such as \lcio and \cernroot.
Recently, the number of HEP-specific developments has seen an uptick, with active developments ranging from statistical analysis tools~\cite{oliver_schulz_2019_3568167}, 
over readers for the \cernroot file format~\cite{unroot_jl,uproot_jl}, to an interface to data from the Particle Data Group~\cite{tamas_gal_2020_3933364}. \julia starts 
being used to prepare data for HEP publications, e.g.~\cite{Tastet:2019nqj}\footnote{Based on private communication with one of the authors.}, and a toolkit for Bayesian analysis~\cite{Schulz:2020ebm} has been implemented in \julia. The main item 
missing from the eco-system for carrying out a PhD-level physics analysis is an implementation of HEP-specific likelihood fitting codes.

Nevertheless, we argue that the ecosystem is mature enough to support a large variety of tasks encountered in HEP, and here we demonstrate a 
couple of workflows. All plots in this paper were created using \julia packages.

This paper is organized as follows: First, we give a brief introduction into the features and capabilities of \julia, then we describe the implementation, the setup, and 
the results of the benchmarks for typical analysis tasks, and then we summarize our results.

\section{The \julia Language}\label{section:julia}

The \julia language is a multi-paradigm, dynamic language with optional typing and garbage collection that achieves good run-time performance by 
using a just-in-time compiler (JIT). The runtime supports distributed parallel as well as multi-threaded execution.
It has been demonstrated to perform at peta-scale on a high-performance computing platform~\cite{2018arXiv180110277R}, and it has strong support for scientific machine 
learning~\cite{2020arXiv200104385R,Flux.jl-2018,innes:2018,Yuret2016KnetB}. 
The language implementation is open source, available under the MIT license. It is available for download for Windows, MacOS X, Linux x86, FreeBSD platforms, 
among others~\cite{juliaplatforms}.
Several universities use the language in their programming courses~\cite{juliaclassroom}.
As we show in the following, it is also very well suited for developing software in large distributed collaborations 
like those for high energy physics collider experiments.

\subsection{Key Features}

As \julia is mainly targeted for scientific applications, it supports arbitrary precision integers and floats using the 
GNU Multiple Precision Arithmetic Library (GMP)~\cite{10.5555/2911024} and the GNU MPFR Library~\cite{10.1145/1236463.1236468}. 
Complex numbers and an accurate treatment of rational numbers are also implemented.
The base mathematical library is extensive and includes a linear algebra package with access to specialized libraries such as 
BLAS~\cite{vandeGeijn2011} or MKL~\cite{intelmkl}.
The language fully supports Unicode, which allows the user to write mathematical formulae 
close to how they would be typeset in a book or paper.

\julia has a builtin package manager, which uses a central repository for registered packages, the so-called \julia registry, but the 
user has the choice to define their own registry, or to add unregistered packages. Packages with binary dependencies can be built and 
distributed for a variety of platforms automatically~\cite{binaryBuilder:github}.
This frees users from concerns about compiler versions or library incompatibilities. 


The three main ways to execute code in \julia are described in the following.
\begin{description}
        \item[\textbf{Scripts}] {Similar to \python, code is written to a text file, which can be executed by the \julia executable.}
        \item[\textbf{REPL}] {The default Read-Eval-Print Loop (REPL) in \julia is a highly customizable command line interface, 
        with command history and tab completion. In addition, the REPL comes with several modes, such as a shell mode 
        to execute commands in the shell of the operating system, a help mode to inspect \julia code and a package mode to 
        update and maintain the installed packages. Packages can modify the REPL to make additional modes available.}
        \item[\textbf{\jupyter notebooks}] {\jupyter notebooks have become popular in data science applications, where they provide a convenient way to 
        combine code input blocks with graphical output in the same interface. \julia is very well supported in the \jupyter ecosystem.}
\end{description}

\section{Implementation details and code distribution}
Our study of using \julia in a typical high energy physics analysis workflow uses the
\lcio event data model and persistence format, which has been developed since 
2003~\cite{Gaede:2003ip}. The code is released under the BSD-3-Clause license and hosted 
on Github~\cite{lcio:github}. The release contains 
the \cplusplus, \fortran, \python, and \go code by default. The generation of \fortran bindings (through C 
method stubs) can be enabled by a 
flag at build time. The \python bindings can be enabled at build time, if \cernroot is 
present. In this case, \cernroot can generate dictionaries for the \lcio classes, 
which are then made available to \python through the \pyroot mechanism. 
Additionally, the distribution contains some convenience functions to make the 
bindings more \textit{pythonic}.

The \julia bindings for \lcio (\texttt{LCIO.jl}~\cite{jan_strube_2020_3986687}) were created by wrapping the \cplusplus classes using the \texttt{CxxWrap.jl} \julia 
module and the corresponding \cplusplus 
library~\cite{cxxwrap:github}.
This module requires declaring each class and method to be exposed on the
\julia side to the wrapper generator, which then automatically creates the boilerplate
code to translate between \julia and \cplusplus. This translation is transparent to the user, and we have 
kept the \lcio function names intact, which means that users can still take advantage of the documentation of the upstream library to learn about functionality.
The workflow for this is similar to that for creating {\texttt{Boost.Python}} bindings.

This method allows fine-grained control over how the bindings should behave on 
the \julia side. For example, it allows us to define a more convenient return 
type for the three- and four-vectors in \lcio, which are bare pointers in 
\cplusplus, but properly-dimensioned arrays in \julia. Additionally, the \julia 
bindings allow the use of collections of properly typed objects, where \lcio only provides collections of pointers to a base class.
This frees the user from having to look up the class type for each of the collections that will be 
used in the analysis and is helpful particularly for new users. While the shortcomings in this example are straightforward 
to address on the \cplusplus side, our work shows that this method for defining the bindings is a convenient 
and powerful way to improve the usability of existing libraries without having to modify the underlying code.

For the deployment, we are using the 
\texttt{BinaryBuilder.jl}~\cite{binaryBuilder:github} 
package to automatically build binary distributions for the \lcio library, as well as 
for the \cplusplus component of the wrapper library. The same goes for the \fastjet library and corresponding glue code. 
Since our package was added to the general \julia registry, 
it can simply be installed with the command \jlinl{add LCIO} in the package manager. This downloads the correct compiled binaries 
for the package, as well as the glue code between \lcio and \julia, for the user's operating system, processor architecture and glibc. 
The user does not need to compile the \cplusplus code or worry about compatibility between different library or compiler versions.

\section{The Test Setup}
\label{section:testsetup}
To demonstrate a simple but realistic workflow, we reconstruct the invariant 
mass of the process \epem $\rightarrow \mathrm{Z}^0 \rightarrow$ \mpmm, i.e.\ production of a $\mathrm{Z}^0$ boson in electron--positron collisions 
and subsequent decay to a pair of muons. We also perform simple event-shape 
and jet clustering analyses on a set of \epem $\rightarrow \mathrm{Z}^0 \rightarrow \mathrm{q\bar{q}}$ events, i.e.\ decay of the $\mathrm{Z}^0$ boson to a pair of quark jets.
The events were generated with the \whizard event generator~\cite{Kilian:2007gr}, their interactions 
with the \sid~\cite{Behnke:2013lya} model were simulated using \geant~\cite{AGOSTINELLI2003250,1610988,ALLISON2016186}, and the particles 
were reconstructed using the particle flow algorithms of the \pandorapfa 
package~\cite{2009NIMPA.611...25T}. Events are stored on a dCache system in the 
\lcio format, which uses zlib for compression. The average file size for the data files is 775 MB; each file contains 32400 events. 
To exclude potential OS variations, all results were run on a CentOS Linux 7.7 installation, with 20 Intel(R) Xeon(R) CPU E5-2640 v4 @ 2.40GHz CPUs and 128 GB RAM.

We have implemented the analysis code in \cplusplus and \python, as well as in \julia and made it available online~\cite{jan_strube_2020_3911414}.
All three implementations share the same underlying routines for reading \lcio files. For the computation of the invariant 
mass in \cplusplus and \python, we use the \texttt{TLorentzVector} class of the \cernroot 
package; this is common practice, particularly among students. For simplicity, 
we are not implementing a vector class with a Minkowski metric in \julia and instead compute the 
invariant mass from energy and momentum explicitly. 

For our tests the following versions were used:\\
\begin{center}
\begin{tabular}{l p{2.5in}} \toprule
\textbf{\cplusplus}: 	& gcc 8.3.0 \\
\textbf{\python}:	& Python 3.7.6 \\
\textbf{\cernroot}:     & ROOT 6.22.0 \\
\textbf{\julia}:  	& Julia 1.5.0 \\ \bottomrule
\end{tabular}
\end{center}

The \cernroot version used by the \cplusplus and \python programs was 6.08/06. The scientific diagrams in this document were 
created using the histogram (\texttt{OnlineStats.jl} and \texttt{Plots.jl})
and fitting packages (\texttt{Distributions.jl}) from the \julia General registry~\cite{jlangRegistry}.

\section{Benchmarking}\label{section:Results}

We compare the performance of different implementations for a number of toy analysis tasks. These tasks are simple, but representative workflows that include an I/O-dominated application, an 
application with a heavy computational component, and an application where the computation is carried out by a third-party library. Finally, we compare implementations of a complete processing chain at different levels of parallelism.

\subsection{Processing the \epem $\rightarrow \mathrm{Z}^0 \rightarrow$ \mpmm sample}

In this simple toy analysis we process a series of \lcio files 
with \epem $\rightarrow \mathrm{Z}^0 \rightarrow$ \mpmm events to 
reconstruct the $\mathrm{Z}^0$ mass using a Gaussian fit. A real-world application would presumably use a more appropriate shape for the fit, but it shall suffice for the purpose of this study. This is mostly an I/O-dominated application. 

\begin{figure}
    \centering
    \includegraphics[width=0.95\textwidth]{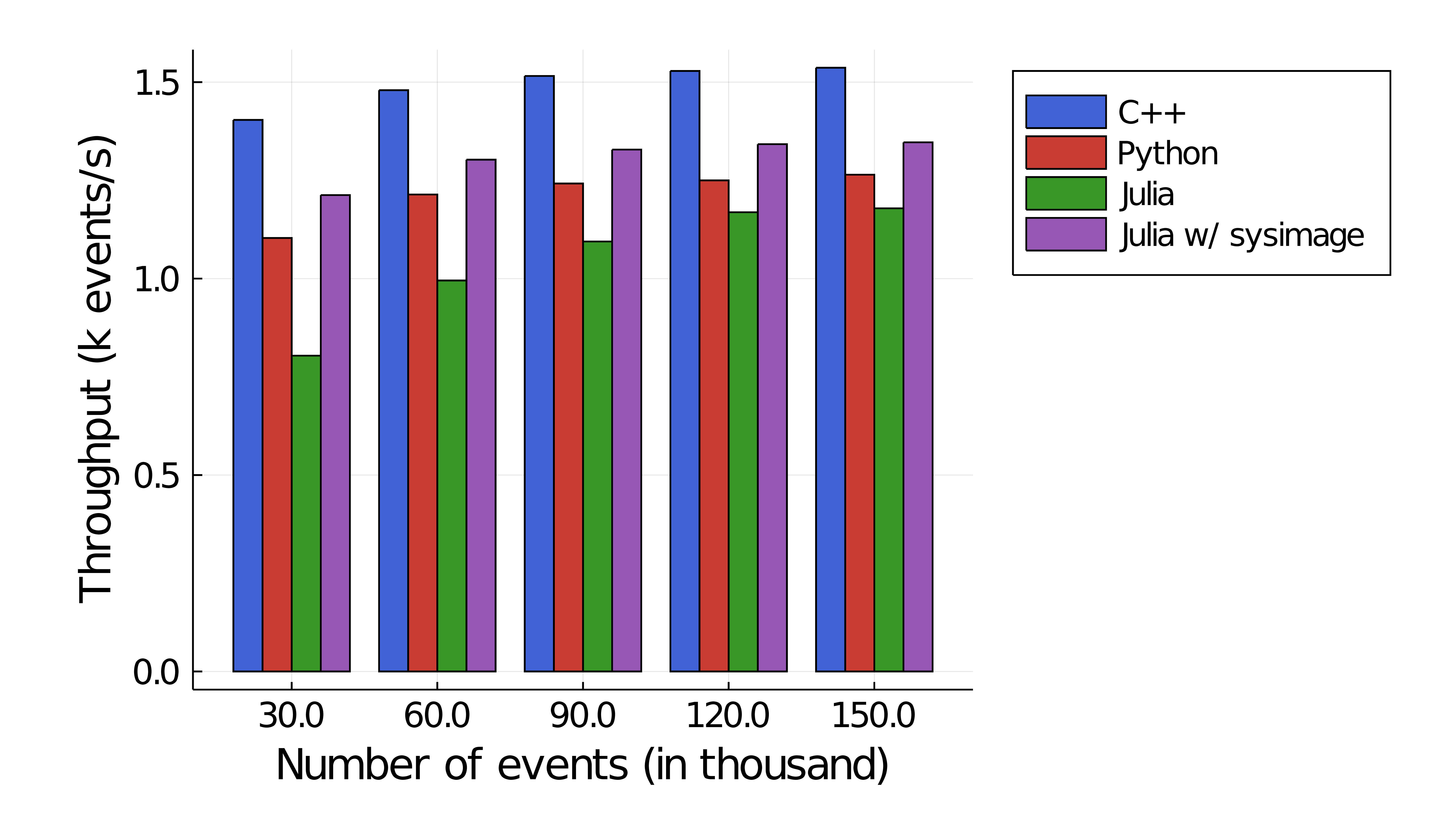}
    \caption{Event throughput in 1000 events per second for reading different sample sizes and fitting the invariant mass of the $\mathrm{Z}^0$ boson.}
    \label{fig:sampleThroughput_grouped}
\end{figure}

For the small-size data set of 30000 events, the \cplusplus implementation is the clear winner, 75\% faster 
than \julia (see Tab.\ref{tab:resultszmumu}). Adding the compile time for the first run of the \cplusplus variant 
of 1.76~seconds does not change the conclusion significantly. This picture changes when one looks at the larger data sets.
The throughputs of the \python and \cplusplus implementations remain more or less constant, as can be expected 
for interpreted and ahead-of-time compiled languages, respectively.
\julia, on the other hand, shows a trend of growing throughput as the data set size increases. 
This behavior can be attributed to the diminishing contribution of the overhead of the just-in-time compilation step.
The trends of processing times for the three language implementations are shown in 
Fig.~\ref{fig:sampleThroughput_grouped}. For comparison, we have compiled the package with the largest start-up time, our 
\texttt{LCIO.jl}, into a static library that the \julia binary links against. This is version is called ``Julia w/ sysimage'' in 
our plots. One can see that the much reduced overhead compared to the default version leads to a nearly constant throughput, 
similar to the \cplusplus and \python versions. We should note, however, that we would expect most users to run the default version with 
the increased start-up cost.

\begin{table}[htbp]
\centering
\caption{Comparison of the performance of \julia, \python and \cplusplus in processing \epem $\rightarrow \mathrm{Z}^0 \rightarrow$ \mpmm events. 
 The total event samples sizes were 30000 and 150000 events, respectively. In order to reduce fluctuations due to external effects like changes in the network or I/O performance, 
 each measurement has been repeated five times. We report here the average and standard deviation of the five runs.}
\label{tab:resultszmumu} 
\begin{tabular}{lccc}
 & \julia & \python & \cplusplus \\
\toprule
Run time (30000 events) (s) & $37.32 \pm 1.24$	&	$27.19\pm 0.55$	& $21.37\pm 0.18$\\
Average number of events / s & 803.88 & 1103.40 & 1403.88 \\
Run time (150000 events) (s) & $127.24\pm 4.24$	&	$118.62\pm 1.86$ & $97.61 \pm 0.34$\\
Average number of events / s & 1178.91 & 1264.50 & 1536.79 \\
Overhead (s) & 15.28 & 3.90 & 2.35 \\
\bottomrule
\end{tabular}
\end{table}

\begin{figure}[htbp]
    \centering
\includegraphics[width=0.95\textwidth]{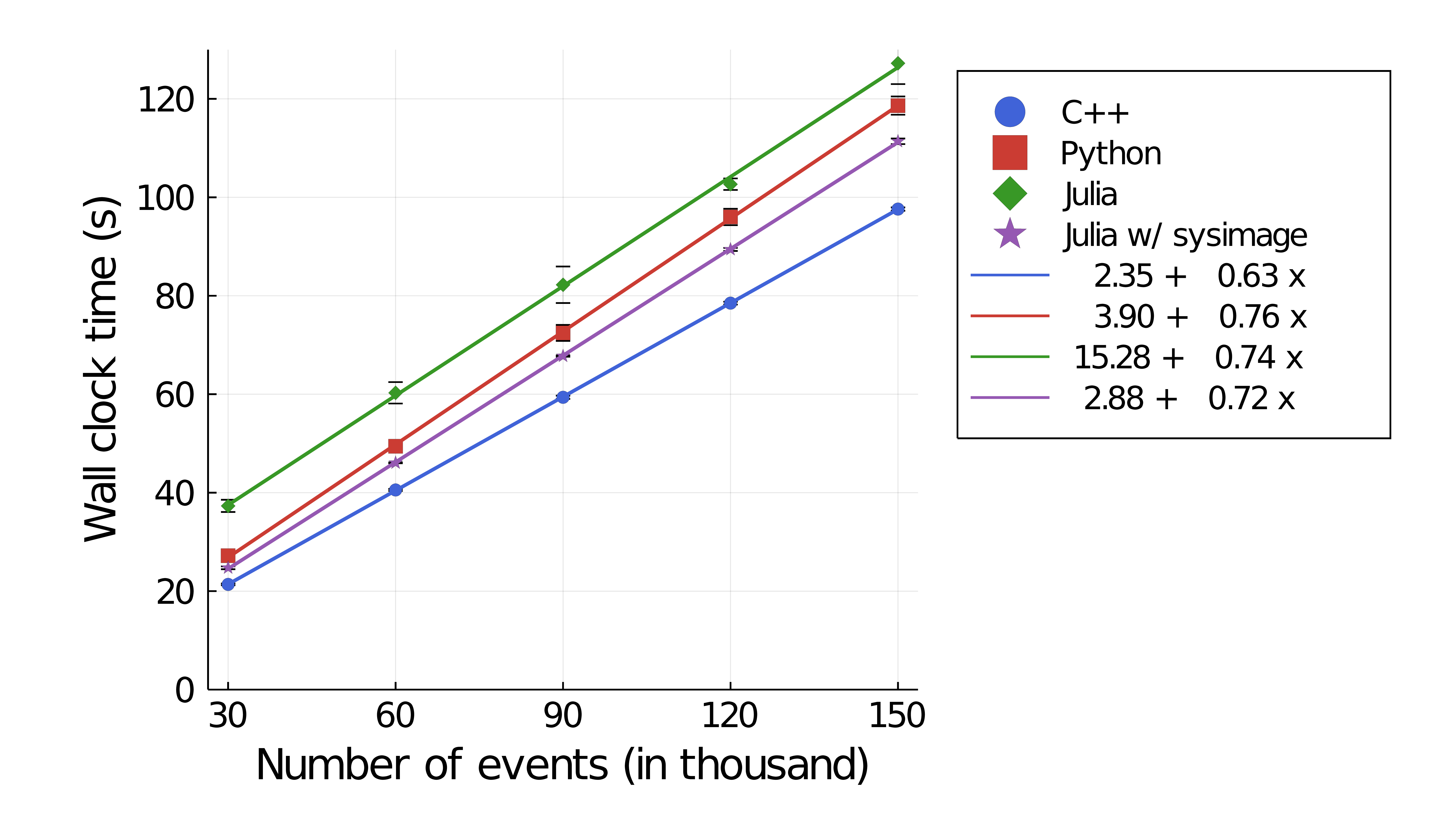}
\caption{The dependence of the execution time on the number of events processed for different implementations.}
\label{fig:zmumu_events_vs_time}
\end{figure}

The question is now how large the overheads of the just-in-time compiled (\julia) and interpreted (\python) languages are 
compared to the \cplusplus implementation.
To get an estimate for this, the execution time (using the Unix \texttt{time} command) 
depending on the number of events processed was measured for the \epem $\rightarrow \mathrm{Z}^0 \rightarrow$ \mpmm events 
data set and fit with a linear regression. The overhead in \julia is clearly visible, giving the largest intercept at 15.28~seconds. 
This is consistent with time measurements within \julia (that start measuring \emph{after} the initialization step), 
which are consistently 15~seconds below what is reported by the Unix \texttt{time} command. For \python, we measure an intercept 
of 3.90~seconds. The intercept for \cplusplus, 2.35~seconds, can be interpreted as the baseline for the system, and is most likely due to 
latency of the access to the data files and libraries hosted on different servers.
The results are summarized in Tab.~\ref{tab:resultszmumu} and Fig.~\ref{fig:zmumu_events_vs_time}. 

\subsection{Event Shape Analysis of Hadronic $\mathrm{Z}^0$ Events}
\label{sec:foxWo}
While studies of an I/O-dominated process are useful tests of a realistic workflow, the advantage of \julia over \python 
comes with a more processing-intense algorithm, possibly with multiple nested loops over reconstructed objects.
As an example of such an algorithm with a realistic use case, classical event-shape variables like thrust and the Fox--Wolfram Moments~\cite{Fox:1978vu} were 
implemented following the implementations in \pythia 6.4 and \pythia 8.2~\cite{Sjostrand:2006za,Sjostrand:2014zea}.
For the analysis a set of \lcio files with \epem $\rightarrow \mathrm{Z}^0 \rightarrow \mathrm{q\bar{q}}$ events was used. 

The implementation in \julia (\texttt{EventShapes.jl}~\cite{jan_strube_2020_3698379}) is a straightforward translation of the \cplusplus code, 
which we have verified to produce identical output. This process revealed another important feature of 
\julia: We found the availability of tools to support the development process to be of a high quality, particularly for identifying the cause 
for slowdowns and sources of memory allocation. While excellent tools exist to identify the sources of segmentation faults in 
\cplusplus, (e.g.\ \texttt{gdb}), and to track memory allocations and benchmark function calls (e.g.\ \texttt{callgrind} and the \texttt{gperf} tool suite), 
several such tools in \julia are built in and part of the downloaded code, and the interactive nature of the language makes it very easy
to isolate pieces of code for benchmarking and debugging. Additional tools are available in third-party packages, and frequently 
it is sufficient to add macros to the code under investigation to learn more about its run-time performance.

Our first version 
was several times slower than the \cplusplus version, but a simple annotation with the \jlinl{@btime} macro from the \jlinl{Benchmarking} 
package helped us realize that this was mainly due to memory allocations in the inner loop. Another frequent cause of 
slowdown is type instability, which prevents the compiler from generating specialized code. In \julia, it is straightforward to 
identify this symptom with the \jlinl{@code_warntype} macro.
While these features aided us in our code translation, they are even more important in cases where no reference code 
exists and implementations of algorithms are developed from scratch. The current version of our \julia implementation 
achieves a processing rate of 287 events per second, while the \cplusplus implementation processes 182 events per second.

Similar to our \julia version, the implementation in \python started out as a straightforward translation of the \cplusplus 
code. It is understood that this does not result in code that is optimal for 
performance, but an important aspect of productivity is how close the performance of a first implementation is to optimal. 
Additionally, it is important that algorithms can be implemented in 
a straightforward manner from a scientific paper. Vandewalle, Kovacevic and 
Vetterli~\cite{4815541} define five levels of reproducibility, with the criteria 
for the highest and most desirable level requiring the research to be 
re-implemented within 15 minutes or less. 

Our naive translation of the \pythia code to \python processed 2000 events in 61~minutes and 14~seconds.
To allow a fairer comparison with the \julia code, we have undertaken some efforts in speeding up the \python code as well.
Our first attempt was to apply an optimization that is common in \numpy applications, namely vectorization across the outer 
dimension of the array (i.e., the innermost loop). In our case, however, this did not lead to a measurable improvement in execution speed.
Our second attempt was to apply the just-in-time compilation of \textsc{numba}~\cite{Numba}. While the recommended way to annotate the 
functions did not work for us, due to its inability to ascertain a specific type for some of the \python objects for the compilation step, 
the \texttt{nopython=True} option led to a dramatic speed-up in our tests, processing 204 events per second. We learned from these tests, 
though, that this comes at the cost of significantly complicating the debugging, because the \python interpreter loses the 
ability to inspect the \textsc{numba}-annotated sections of the code.
Further optimizations are certainly possible, for example by implementing the code in \textsc{cython}~\cite{cython} or by including tools from the 
\scipy~\cite{2020SciPy-NMeth} distribution.

\subsection{Jet Clustering}
We anticipate that most users will use \julia for data analysis in high energy physics in two ways: Either, by writing code directly in 
\julia, for example, by translating existing codes, as demonstrated in Section~\ref{sec:foxWo}, or by calling into existing \cplusplus 
libraries.\footnote{Interacting with libraries written in \python is straightforward in \julia, through the \texttt{PyCall.jl} package.} 
As a demonstration of the latter, we have written simple bindings to the 
frequently used \fastjet package (\texttt{FastJet.jl}~\cite{jan_strube_2020_3929866}) in \julia, using the same 
\texttt{CxxWrap.jl} package that we used to create the \lcio bindings. Similarly to the \lcio bindings, the code has been added to the 
\julia registry and can be installed with \texttt{add FastJet} in the \julia package manager.
Processing the same sample as in Section~\ref{sec:foxWo}, we achieved 319 events per second for the \julia implementation 
and 445 events per second for the \cplusplus implementation. 

\subsection{Parallelizing in \julia}
As the number of cores on modern processors keeps growing, the \emph{event-level parallelism} that has been exploited by 
high energy physics experiments for decades is no longer sufficient to take optimal advantage of the available processing power, 
mainly due to the memory required to process a single event. \julia 
has several constructs to support parallel programming. For the purpose of this section, and following the \julia documentation, 
we distinguish here between \emph{asynchronous}, \emph{distributed}, and \emph{multi-threaded} programming.

In \emph{asynchronous} programming, different pieces of the code run independently of each other, and a scheduler takes care of assigning processing 
cycles to them. For example, a program could read a file concurrently with, i.e.\ at the same time as, setting up a canvas for plotting, since the I/O 
usually has a small ratio of CPU time over wall clock time. This is a straightforward way to speed-up programs that wait for I/O tasks to complete, 
and this level of concurrency is also available in \python, e.g.\ in the \texttt{asyncio} module. \cplusplus supports this level of programming as 
\emph{coroutines} in the \cplusplus 2020 standard.

In \julia, the \jlinl{Distributed} module 
allows scheduling tasks in different processes, either on the same CPU, or on different CPUs that are connected by a network.
In a \emph{distributed} application, different pieces of the code run in different processes. Similar facilities are available in \python 
(for use on the same node) in the \texttt{multiprocessing} module. We are not aware of an implementation in the \cplusplus standard library, 
but a commonly used library for using distributed computing across different computers is \emph{MPI}, while shared-memory parallelism can be 
implemented by using programming interfaces like OpenMP and TBB~\cite{inteltbb}.

The level of thread-\emph{multi-threaded} programming that is supported in \julia since version 1.3.0 is not available in \python, due to 
the global interpreter lock (GIL). In this level of parallelism, different parts of the program run in the same process space. On multiple 
cores, they can be scheduled such that each thread takes full advantage of a different core. \cplusplus has support for using different threads 
since the addition of the \emph{threads} library to the \texttt{C++11} standard.

Our implementation of multi-threaded event processing is based on \julia \jlinl{Tasks} that we spawn on different threads, and we have 
implemented parallel programming concepts on two levels. 
The communication between different \jlinl{Tasks} happens via \jlinl{Channels}, which promotes a design similar to that of the 
\go programming language, for example.
We are combining data readers on separate \emph{distributed} processes (in our case on the same CPU) using the \jlinl{Distributed} module
with data processors running in parallel on different threads using the \jlinl{Threading} module. Combining these two levels in \julia 
is straightforward, as shown below. This allows us to make optimal use of the available processing power given the specific I/O and processing 
characteristics of our application.

\subsubsection{File-parallel Data Access}
Many \cplusplus libraries used in high energy physics analyses are not thread-safe. While \lcio has recently gained a thread-safe mode of operation, we are using the conventional, 
non-thread-safe version of the library to demonstrate a code pattern that allows using such \cplusplus libraries in a parallel program nevertheless. For this, we are taking 
advantage of the fact that common workflows operate on data sets that frequently span multiple files. In our implementation, we combine several 
different \lcio readers, each running in its own independent process and reading a different set of files. This comes at the cost of having additional overhead from inter-process communication, but our 
approach is applicable to basically any \cplusplus library, regardless of whether it is thread-safe or not. Furthermore, it is straightforward to change 
the number of concurrently running instances of \lcio to optimize the ratio of event readers to event processors. In our implementation, 
the different \lcio processes run on the same node, but the extension of this kind of concurrency to multiple connected nodes is straightforward. 
This would allow processing parts of the data set hosted on different machines, but it comes with its own trade-offs of network connectivity, 
communication between the nodes, and the processing power present on the individual nodes. A detailed investigation of such a use case exceeds the scope of this paper. 

\subsubsection{Putting It All Together: Processing Event Shapes and Jets in Parallel}
Since \julia version 1.3.0, the runtime supports multi-threaded execution, which simplifies the implementation of \emph{within-event parallelism} tremendously. 
To demonstrate the achievable speed-up, we are processing the event thrust and the Fox--Wolfram moments, implemented in \julia, and the jet clustering in \cplusplus \fastjet, 
for each event. Our example executes multiple event analyzers in parallel on different threads. While this paradigm allows for shared-memory parallelism, we have opted here 
to also use \jlinl{Channel}s for communication. 
In the following we sketch the implementation in \julia, show-casing how straightforward it is to set up multi-threaded event processing 
using the \jlinl{Distributed} package.
The number of available \texttt{workers} can be defined during the \julia startup, e.g. \jlinl{julia -p 8} would start eight worker nodes within \julia. 
However the number of threads needs to be set using the environment variable \texttt{JULIA\_NUM\_THREADS} before or via the \texttt{-t} flag when 
invoking the \julia executable.

In the \julia main function, several \jlinl{RemoteChannel(()->Channel())} are being opened to ensure communication between the data readers and the data processors. The channels 
are buffered to a given depth, which allows the readers to continue filling them up to the buffer depth, while the processors deplete them from the other end. A write on a full 
buffer blocks until at least one element has been removed from the other end.
The code skeleton of the \texttt{main} function for this is shown in Listing~\ref{listing:mt:main}, illustrating the usage of channels and spawning the data readers 
on specific processes (using \jlinl{@spawnat}), and the spawning of the data processors on different threads within the main process using the \jlinl{Threads.@spawn} command.

\begin{lstlisting}[caption=The \julia main function used for the parallel event processing,label=listing:mt:main]
function main()
	# let's start with reading up to four files concurrently
	fnames = RemoteChannel(()->Channel{String}(4))

    # let's presume we can buffer up to 200 events.
    # Adjust according to available memory and 
    # relative speed of readers and processors.
	events = RemoteChannel(()->Channel(200))

	# the data readers can signal when they are done reading events
	done = RemoteChannel(()->Channel{Int}(4))

	# up to 10 data processors 
	nProcessed = RemoteChannel(()->Channel{Int}(10))

	# spawn the data readers
	for w in workers()[1:NREADERS]
		@spawnat w readEvents(fnames, events, done)
	end	
	# spawn the data processors, one per thread on this process
    processors = [
        Threads.@spawn processEvents(events, nProcessed) 
        for w in 2:Threads.nthreads()
    ]

    # we now have several functions waiting for input on their channels
    # prepare the file names
    for f in ARGS
            put!(fnames, f)
    end
    close(fnames)

    # wait for all readers to be done
    # then we can close the event queue and the processors can finish
    nDone = 0
    nEvents = 0
    while nDone != nworkers()
            nEvents += take!(done)
            nDone += 1
    end
    close(events)

    # wait for the processing to finish
    totalProcessed = 0
    for p in processors
            wait(p)
            totalProcessed += take!(nProcessed)
    end
    close(nProcessed) 
    # if all of the channels are closed, the Tasks can finish
    # at this point, we should have: nEvents == totalProcessed
end
\end{lstlisting}
	
The data reader function is shown in Listing~\ref{listing:mt:reader}. The \jlinl{@everywhere} macro instructs \julia to make this function available on all processes.
The data reader ingests events from an \lcio file and pushes them into a \jlinl{RemoteChannel()}. 

\begin{lstlisting}[caption=The data reader function used for the parallel event processing,label=listing:mt:reader]
@everywhere function readEvents(fnames, events, done)
    iEvents = 0
	while true
        try
            # take the next file out of a Julia channel
			fn = take!(fnames)   
			LCIO.open(fn) do reader
				for evt in reader
				    # write the read Event into the event channel
                    put!(events, collection) 
                    iEvents += 1
				end
			end
		catch e
			break
		end
    end		
    # write the number of processed events to done channel
	put!(done, iEvents) 
end
\end{lstlisting}

The data processors that have been spawned using \jlinl{Threads.@spawn} are now listening on the \jlinl{RemoteChannel()} for events being 
available for processing as shown in Listing~\ref{listing:mt:processor}.

\begin{lstlisting}[caption=The data processor function used to demonstrate multi-threaded processing,label=listing:mt:processor]
function processEvents(events, nProcessed)
	while true
		try
			collection = take!(events) #receive next event
			h10, h20, h30, h40 = EventShapes.foxWolframMoments(collection)
			#more analysis and clustering
		catch e
			#store number of Events that have been processed
			put!(nProcessed, iEvents) 
			break
		end
	end
end
\end{lstlisting}

\begin{figure}
    \centering
    \includegraphics[width=.975\linewidth]{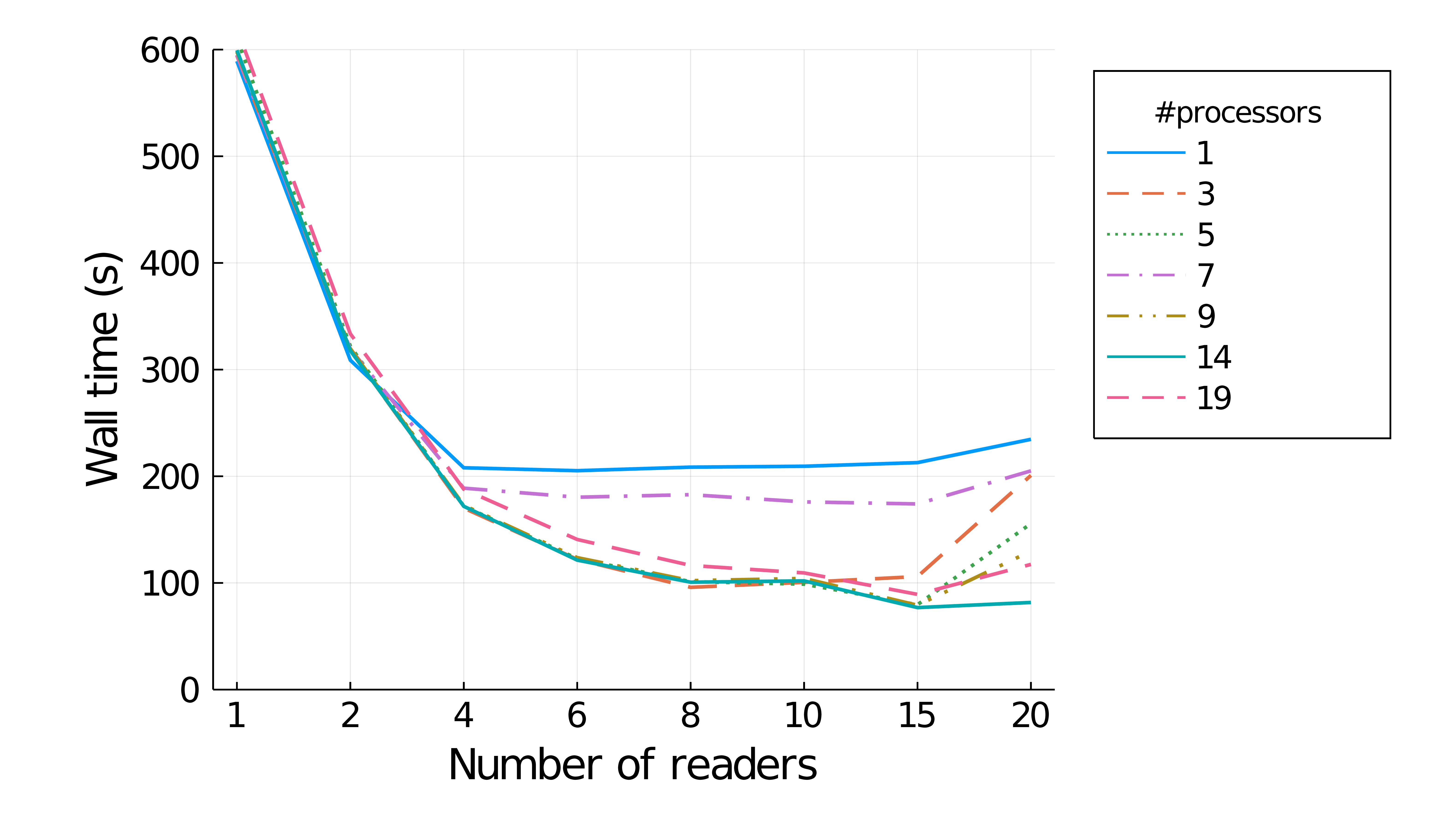}
    \caption{Time (in seconds) for processing 233857 events in 23 files with different numbers of data readers and data processors in \julia.}
    \label{fig:threading_speedup}
\end{figure}

The available speed-up using this paradigm is shown in Fig.~\ref{fig:threading_speedup}. It shows 
clearly that the processing is I/O--dominated: More data readers give a better performance, up to a maximum of 15. The maximum value of 20 readers in our test leads 
to a decrease in performance, most likely due to 
contention between different processes. Additionally, one can see that more data processors do not necessarily lead to a better overall throughput. The optimal values for numbers of readers and processors, as well as the 
buffer depth on the \texttt{Channel} between the two, depend on the workload and the details of the hardware configuration. A detailed analysis exceeds the scope of this 
paper, but our chosen paradigm for setting up concurrent processing makes it trivial 
to evaluate the performance for a given setup and optimize parameters for larger processing jobs.

\begin{figure}
    \centering
    \includegraphics[width=.975\linewidth]{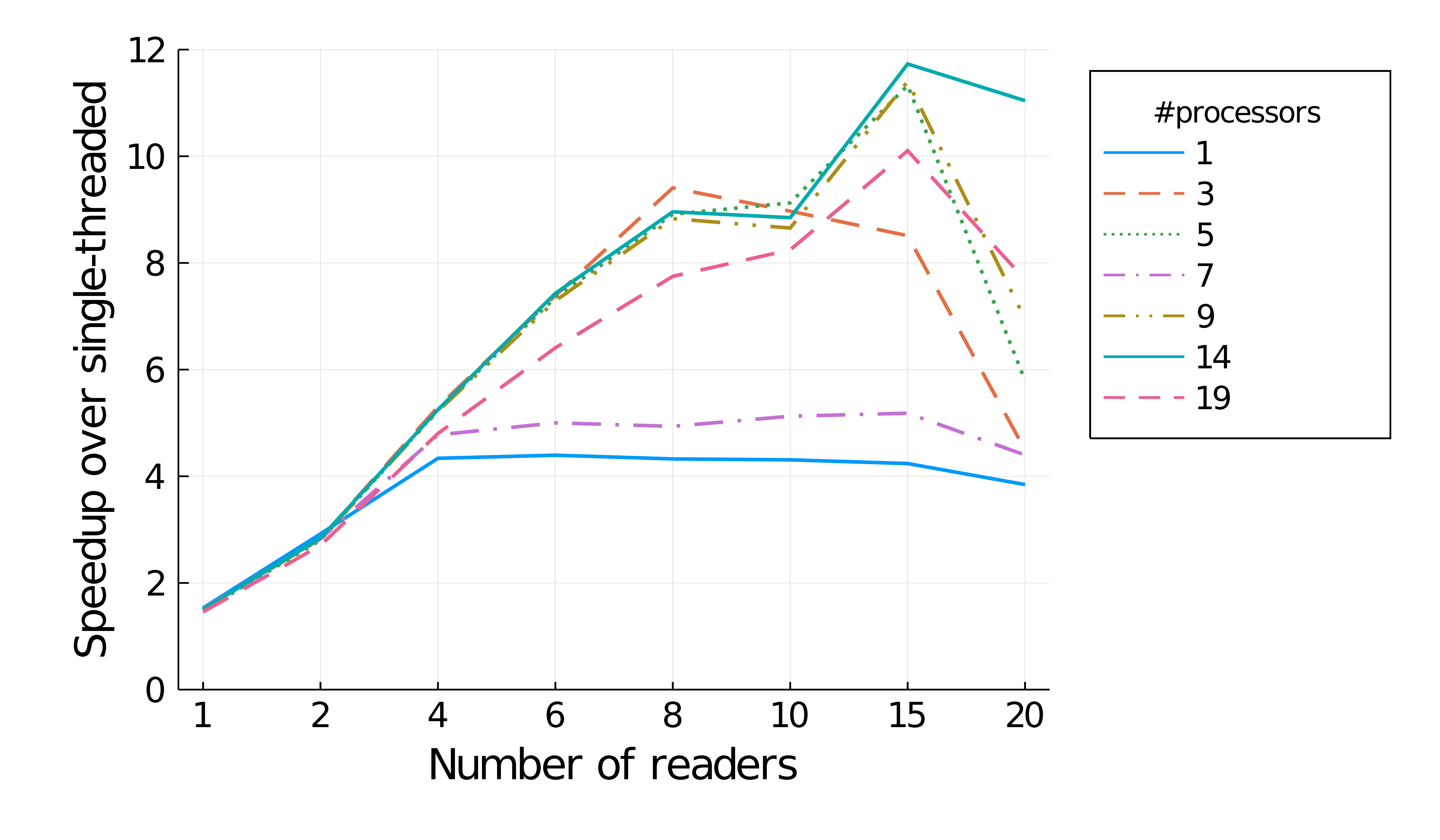}
    \caption{Relative speed-up of the multi-threaded \julia code relative to the single-threaded application for different numbers of concurrent data readers and data processors. The single-threaded application processes 259 events/s.}
    \label{fig:speedupOverCPP}
\end{figure}

Fig.~\ref{fig:speedupOverCPP} shows the relative speed-up that can be achieved in this way over a single-threaded \cplusplus application.
For reference, a single-threaded application processing the events in \julia takes 15 minutes and 2 seconds and the \cplusplus processing takes 22 minutes and 25 seconds.

\section{Summary}\label{section:Summary}
For many years, the high energy physics community has spent significant 
resources on adding interactivity to their software to facilitate plotting and data exploration.
Data science packages developed by other fields have realized the same need for interactive data access, and the 
standard packages in this sector are all accessible from the interpreted language \python.
In this paper, we have demonstrated that the \julia programming language can offer a compelling option for data analysis in high energy physics.
The interactivity of the language is on par with \python, with support for \jupyter 
notebooks and a rich ecosystem for plotting and statistical data analysis, as well as deep learning. 

With a large set of specialized codes in \cplusplus and \fortran, developed over 
several decades, interoperability with these languages is a key requirement for 
any data analysis solution in high energy physics. We have shown that it is 
straightforward to interface \julia with existing libraries. What is more, the 
package manager in \julia makes it easy to install the code, without requiring the availability of a compiler for the other languages.
The overhead of calling these third-party codes is measurable for simple applications, but negligible for applications with a computation-heavy component.

Fast execution is a requirement for code that has to process millions 
or even billions of collision events. We have demonstrated that complex 
algorithms consisting of loops nested multiple levels deep, when implemented in \julia, perform on 
par with, or even better than, an implementation in \cplusplus. The language and package ecosystem 
have strong support for debugging and studying the software performance. As the language has reached version 1.0 only two years ago, we expect that many opportunities for optimizing the performance further can still be exploited.

The increased use of many-core systems and the growing demand on multi-threading applications to take advantage of the available hardware position 
\julia extremely well for a growing role in scientific applications in general. We have demonstrated a straightforward way to speed up processing of 
existing algorithms, without requiring that the algorithms themselves be multi-threaded. The composable parallelism in \julia allows combining this 
level of parallelism with algorithms that are themselves multi-threaded in nature\footnote{For a more detailed discourse on this topic, the interested 
reader is referred to \url{https://julialang.org/blog/2019/07/multithreading/}}. A growing set of examples of such algorithms are available online.
This feature is what makes \julia an excellent choice as an analysis language in high energy physics, as it allows exploring new algorithms directly
at the analysis level without having to drop down to the underlying \cplusplus reconstruction framework for want of better performance.

In summary, \julia currently presents a compelling option for data analysis in 
high energy physics, and its multi-threading capabilities positions \julia well for future developments. 
A common challenge for supervisors of summer students is which language to 
teach them. With \cplusplus, significant time is spent teaching them the details 
of memory management and how to avoid segmentation faults. This is particularly true when interfacing with analysis code in 
high energy physics, much of which is written in \texttt{C++98} style and looks very different from the recommended syntax in \texttt{C++17} or newer.
With \python, on the other hand,
different packages use different syntax to avoid the inherent slowness of 
\texttt{for} loops. \julia advertises itself as solving the two-language 
problem, where users combine a statically compiled language for 
performance-critical code with an interpreted language for interactivity. Our 
studies show that the language keeps this promise and it is straightforward to write high-performance code.
The language enables an interactive exploration of data and facilitates the exploration of complex algorithms and is thus an excellent partner
to the \cplusplus processing frameworks used in high energy physics.

\begin{acknowledgements}
    The authors gratefully acknowledge the feedback and support of the \julia community, in particular Bart Janssens, the author of \texttt{CxxWrap.jl}.
    We acknowledge DESY (Hamburg, Germany), a member of the Helmholtz Association HGF, for the provision of experimental facilities. Parts of this research were carried out at the National Analysis Facility (NAF). 
    MS acknowledges the support from DESY (Hamburg, Germany), a member of the Helmholtz Association HGF.
\end{acknowledgements}

\section{Conflict Of Interest Statement}
On behalf of all authors, the corresponding author states that there is no conflict of interest.

\bibliographystyle{spphys}       
\bibliography{bibliography}   

\end{document}